\documentclass[fleqn,twoside]{article}
\usepackage{espcrc2}

\usepackage{graphicx}
\usepackage[figuresright]{rotating}

\newcommand{\AmS}{{\protect\the\textfont2
  A\kern-.1667em\lower.5ex\hbox{M}\kern-.125emS}}

\newcommand{\be} {\begin{equation}}
\newcommand{\ee} {\end{equation}}
\newcommand{\GeV}{\mbox{\rm GeV}}

\title{$B_K$ from improved staggered quarks 
\thanks{Talk presented by E. G\'amiz at
        {\it Lattice 2004}, Fermilab, 21--26 June 2004.} }

\author{E.G\'amiz\address{Department of Physics \&
               Astronomy, University of Glasgow, Glasgow, G12 8QQ, UK.},
        S.Collins$^{\rm a}$,
        C.T.H.Davies$^{\rm a}$, 
        J.Shigemitsu\address{Physics Department, The Ohio State
        University, Columbus, OH 43210, USA.}, 
        M.Wingate\address{Institute for Nuclear Theory, University of
        Washington, Seattle, WA 98115, USA.}, 
        HPQCD and UKQCD collaborations.
       }
\begin{document}

\begin{abstract}
We compare calculations of $B_K$ with improved staggered quarks (HYP, Asqtad) 
and demonstrate the improved scaling behaviour that this gives rise to 
over previous calculations with unimproved staggered quarks. 
This enables us to perform the calculation of $B_K$ 
on the MILC dynamical configurations ($n_f=2+1$), for which we give 
preliminary results.  
\vspace{1pc}
\end{abstract}

\maketitle

\section{Introduction}

Indirect CP violation in the system of neutral kaons is parametrized by 
$\varepsilon_K$. Experimentally it is a very well measured quantity. On the 
other hand, its theoretical expression depends on certain CKM matrix elements 
about which we would like to obtain information and on the hadronic matrix 
element between $K^0$ and $\bar K^0$ of the effective hamiltonian describing 
processes with $\Delta S=2$, 
\be 
H_{eff}^{\Delta S=2}=C_{\Delta S=2}(\mu)\int d^4x \,Q_{\Delta S=2}(x)
\ee
with
\be
Q_{\Delta S=2}(x) = \left[\bar s_{\alpha}\gamma_{\mu}
d_{\alpha}\right]_{V-A}(x)
\left[\bar s_{\beta}\gamma^{\mu}d_{\beta}\right]_{V-A}(x).
\ee

The matrix element of this operator is usually 
normalized by its VIA value, defining $B_K$ as the ratio 
\be
B_K(\mu) \equiv \frac{\langle\bar K^0|
Q_{\Delta S=2}(\mu)|K^0\rangle}
{\frac{8}{3}\langle \bar K^0|\bar s\gamma_{\mu}\gamma_5d|0\rangle
\langle 0|\bar s\gamma_{\mu}\gamma_5d|K^0\rangle}.
\ee

The theoretical error associated with the calculation of $B_K$ is the main 
source of uncertainty when one tries to use the experimental value of 
$\varepsilon_K$ to constraint the CKM matrix, so improvement in the 
determination of this parameter is crucial to gain information about the 
unitarity triangle. The lattice calculations have an important role to play 
in such reduction of errors.

\section{Calculation of $B_K$ using staggered fermions}

The staggered quark formulation has   
the advantage of conserving some chiral symmetry at nonzero lattice 
spacing.  This residual symmetry forbids 
the mixing with operators of different chirality leading to a much 
simpler renormalization process. In addition, 
dynamical calculations with staggered quarks are feasible with present 
computers.

The value of $B_K$ used by phenomenologists at 
present in CKM studies was obtained by the JLQCD collaboration 
\cite{JLQCD97} using unimproved staggered quarks in the quenched approximation. 
The value given in \cite{JLQCD97} is 
$B_K^{NDR}(2~\GeV)=0.628(42)$. In spite of being 
considered the benchmark of the calculations of $B_K$, the determination 
by the JLQCD collaboration has several drawbacks. First, it presents 
large scaling corrections, so the extrapolation to the continuum is not 
as reliable as it could be with a better scaling behaviour. 
The main uncertainty is quenching, whose possible impact was estimated 
in \cite{Sharpe97} using chiral perturbation theory to be of the 
order of 15\%. The $SU(3)$ breaking effects are not  
incorporated in the calculation either, but this probably has a minor effect 
\cite{Sharpe97}. We show here that the large scaling violation 
can be corrected by using improved staggered fermions. On the 
other hand, the dominant error, associated to quenching, can be reduced by 
performing dynamical simulations. And, finally, the use of kaons made of 
two non-degenerate quarks can account for the $SU(3)$ breaking effects. 
This three-fold improvement of the staggered determination is the goal 
of our calculation.

\section{Improved staggered actions}

\begin{figure}
\includegraphics[angle=-90,width=75mm]{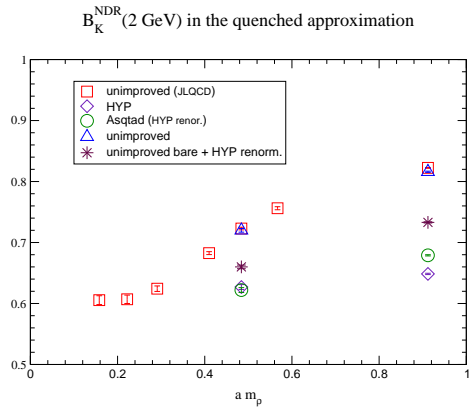}
\caption{Scaling of $B_K^{NDR}(2\GeV)$ with $a$ for improved staggered actions 
compared to the JLQCD 
unimproved staggered results. The stars are values calculated by 
renormalizing the unimproved bare matrix with the HYP Wilson coefficients. 
See text for a discussion of renormalization.}
\label{fig:quenched}
\end{figure}

We first check the impact of using improved actions in the quenched 
approximation. We study two such actions, 
the HYP action \cite{HYP} and the Asqtad action \cite{Asqtad}.  
Fig. \ref{fig:quenched} shows our results for unimproved staggered, Asqtad 
and HYP actions along with the JLQCD results, for several values 
of the lattice spacing.

Two comments are in order in relation with 
Fig. \ref{fig:quenched}. 
First of all, we match kaon masses at a given $\beta$ 
to those of the JLQCD collaboration to make a clear comparison. 
Secondly, the Wilson coefficients needed in the 
calculation of the renormalized $B_K$ with the Asqtad action are not 
available at present \cite{renormAsqtad}. Hence, as a first estimate, 
we use the Wilson coefficients for HYP \cite{LS03} in the Asqtad case 
in Fig. \ref{fig:quenched}.  
This is incorrect, but large differences are unlikely because the 1-gluon 
vertex is the same for HYP and the $\overline{Fat7}$ smearing which goes 
into Asqtad.

Fig.\ref{fig:quenched} shows that 
discretization errors are much smaller with the improved actions. This 
improvement is largely due to the change in renormalization factors, 
not to a change in the bare matrix elements. The contribution of several 
matrix elements confuses the picture but one way to show this is 
given in Fig. \ref{fig:quenched}. There, the stars correspond to the values  
of $B_K^{NDR}(2\GeV)$ obtained with the 
unimproved bare matrix elements and the HYP renormalization coefficients. 

We expect improved scaling 
to survive unquenching and therefore allow us to perform 
reliable dynamical 
calculations with only a few values of the lattice spacing and even 
obtain useful information from a single point simulation.

\section{Dynamical effects in the calculation of $B_K$}

\begin{figure}
\includegraphics[angle=-90,width=75mm]{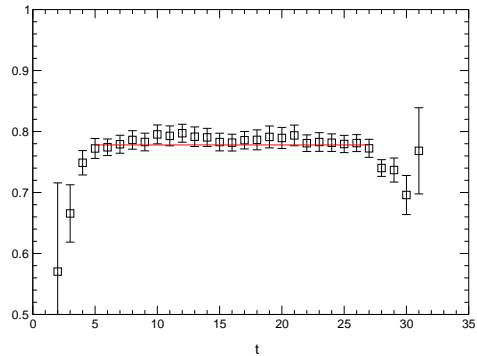}
\caption{$B_K^{bare}$ as a function of the timeslice 
($m_{sea}=0.01/0.05$).}
\label{fig:bkbare}
\end{figure}

The next step is to incorporate dynamical effects. 
We performed a dynamical calculation of $B_K$ within the Asqtad action, 
using the configurations from 
the MILC collaboration with $n_f=2+1$ dynamical flavours \cite{MILC01}. 
Two ensembles at $a=0.125\,fm$ are used: $m_{sea}=0.01/0.05$ and 
$m_{sea}=0.02/0.05$. From the hadron masses determined after the simulation, 
the strange quark mass, $am_s$, was found to be $0.04$. 
We make kaons on the configurations from degenerate quarks of bare mass 
$m_s/2=0.02$.

Fig. \ref{fig:bkbare} shows an excellent plateau when $B_K^{bare}$ is 
plotted as a function of the lattice timeslice.

\begin{figure}
\includegraphics[angle=-90,width=75mm]{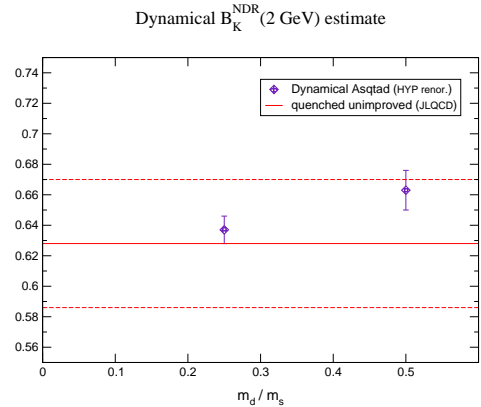}
\caption{An estimate of $B_K^{NDR}(2\GeV)$ from dynamical lattice QCD. 
The x axis corresponds to the light sea quark mass divided by the (real) 
strange quark mass. The lines represent the quenched result from 
\cite{JLQCD97}.}
\label{fig:results}
\end{figure}

In the absence of the correct renormalization coefficients, we again 
estimate $B_K^{NDR}(2\GeV)$ from the HYP Wilson coefficients 
for unimproved glue. 
The results we obtained in this way for the two 
different values of the light 
quark mass, including only statistical errors, 
together with the result quoted by the JLQCD collaboration and its  
corresponding uncertainty, are depicted in Fig. \ref{fig:results}. 

From the results in Fig. \ref{fig:results} 
it seems likely that $B_K$ from full QCD will be close 
to the quenched result. There is also some sign that $B_K^{NDR}(2\GeV)$ 
is falling with the dynamical quark mass.

\section{Summary}

We found that the large scaling corrections that affected previous staggered 
calculations of $B_K$ are reduced 
by using improved actions (HYP, Asqtad) in the quenched approximation. 
This improved behaviour allows us to perform a reliable dynamical calculation 
of this non-perturbative parameter on the MILC configurations. We give 
a preliminary estimate for $B_K^{NDR}(2\GeV)$ in Fig. \ref{fig:results}.

In order to get a final dynamical value for $B_K^{NDR}(2\GeV)$ and 
to clear up the role of dynamical effects in this quantity, we must 
redo our calculation using the correct renormalization coefficients, 
obtain results for another lattice spacing 
and eventually, although its effect is less important, consider physical 
kaons instead of kaons made up of two degenerate quarks with $m_s/2$.

\vspace{.1in}
\noindent
Acknowledgements: E.G. is indebted to the European Union for a Marie 
Curie Intra-European Fellowship. The work is also supported by PPARC and DoE.

\end{document}